# Equilibrium kink-like torsion deformation of a magnetoactive elastomer under a magnetic field

Yu.I. Dzhezherya[1,2,3], A.V. Kyryliuk[1,2], S.V. Cherepov[1], Yu.B. Skirta[1,2], S.O. Reshetniak[1,2,*], S.M. Ryabchenko[3], V.M. Kalita[1,2,3]

[1]V.G. Baryakhtar Institute of Magnetism of the National Academy of Sciences of Ukraine, 36-b, Akademika Vernads′koho Blvd., Kyiv 03142, Ukraine

[2]National Technical University of Ukraine "Igor Sikorsky Kyiv Polytechnic Institute", 37 Beresteiskyi Ave., Kyiv, 03056, Ukraine

[3]Institute of Physics, NAS of Ukraine, 46 Prospekt Nauky, Kyiv 03028, Ukraine

A novel effect involving the formation of a stable kink-like torsion deformation in a magnetoactive elastomer (MAE) beam subjected to a uniform magnetic field is theoretically predicted and experimentally confirmed. The phenomenon was demonstrated using an elastomer beam containing soft magnetic carbonyl iron microparticles within a silicone matrix. The torsion kink acts as a transition boundary between two undeformed homogeneous states of the beam. We show that the elastic moment is compensated by a magnetoelastic moment in the kink region, where the local magnetization of the beam is non-collinear to the applied magnetic field due to shape anisotropy. It is established that within the kink region, the MAE beam exists in a non-uniformly elastically deformed, low-symmetry magnetic state. Outside the kink, the beam's magnetization is collinear with the magnetic field, corresponding to an undeformed, high-symmetry homogeneous magnetic state.

*Corresponding author

E-mail: s.reshetnyak@kpi.ua (S.O. Reshetniak)

Magnetoactive elastomers (MAEs) are composite materials comprised of ferromagnetic particles dispersed within an elastomer matrix[1-5]. Due to the high elasticity of the matrix, MAEs exhibit easy and reversible deformation (stretching, compression, and bending). Significant deformations in MAE samples can be induced not only by external mechanical stress but also by an external magnetic field[6-8], which is of practical interest for creating magnetically controlled contactless devices.

MAEs belong to the class of "smart materials"[9] and hold promise for various applications. They exhibit properties atypical of conventional composites. In MAEs with an elastically soft matrix, a large magnetorheological effect is observed, where the value of the elastic moduli depends on the applied magnetic field strength[10-14]. For soft-matrix samples, the change in elastic modulus values can reach hundreds or even thousands of percent in a magnetic field. This anomalous magnetorheological effect is generally associated with the restructuring of the magnetoactive elastomer, where the spatial distribution of magnetic particles within the matrix changes under the magnetic field's influence, leading to the formation of a chain-like (columnar) structure of magnetized particles[15-18].

The study of MAE properties is of significant scientific interest. Besides the anomalous magnetorheological effect and giant magnetostriction[19,20], MAEs exhibit the magnetic shape memory effect[21,22], the magnetoelectric effect[23], and specific critical deformation effects, such as wave-like sample shape deformation[24] or the critical bending of an MAE beam in a magnetic field[25-27].

The magnetic field-induced critical bending of an MAE beam in a transverse magnetic field is a symmetry effect[28]. At the critical field, the beam's magnetic symmetry spontaneously changes, its magnetization spontaneously deviates from the magnetic field direction, and it leads to the creation of a magnetoelastic moment that bends the beam, thereby changing its shape symmetry. Owing to the specific magnetoelastic properties of MAEs, they are promising for applications as functional materials for origami structures, soft robotics, and metamaterials[27,29-33].

In this work, we investigate the influence of the magnetoelastic moment of forces on the torsional deformation of an MAE beam. We have found that the moment of forces generated by the magnetic field in the magnetized MAE can induce a localized kink-like torsion deformation in the MAE beam, concentrated typically near the centre and largely independent of the beam's total length. The MAE beam with the torsion kink is effectively divided into sections: a part that remains in a high-symmetry magnetic state with magnetization collinear to the magnetic field, and the kinked part, which represents a non-collinear, low-symmetry non-uniform magnetic state. The low-symmetry torsion kink serves as a transition region between the undeformed magneto-homogeneous states of the MAE beam.

Note that in the existing literature, torsional deformation is primarily utilized as a method to determine the shear modulus value of MAEs[34-37]. Experiments typically involve cylindrical MAE samples, where the flat surface is subjected to a torsional mechanical load. Measurements are performed at small deformations, and the magnet is oriented along the axis of the cylindrical samples. In these experiments, the diameters of the samples are significantly larger than their height, so the formation of a torsion kink is impossible. Furthermore, in studies focused on the torsional deformations of MAEs, the magnetic field-stabilized torsion kink has not been investigated, either theoretically or experimentally.

Figure 1(a) shows an image of a thin, long MAE beam, where the height is much greater than the width, which in turn is much greater than the thickness, $a>>b>>c$. The inset shows an enlarged fragment of the MAE beam with magnetic non-coercive particles embedded in the elastomer matrix, where the particle sizes are much smaller than the beam thickness. We assume that the filler particles are spherical, soft magnetic, possess negligibly small magnetocrystalline anisotropy, and are equiprobably distributed throughout the sample/beam volume.

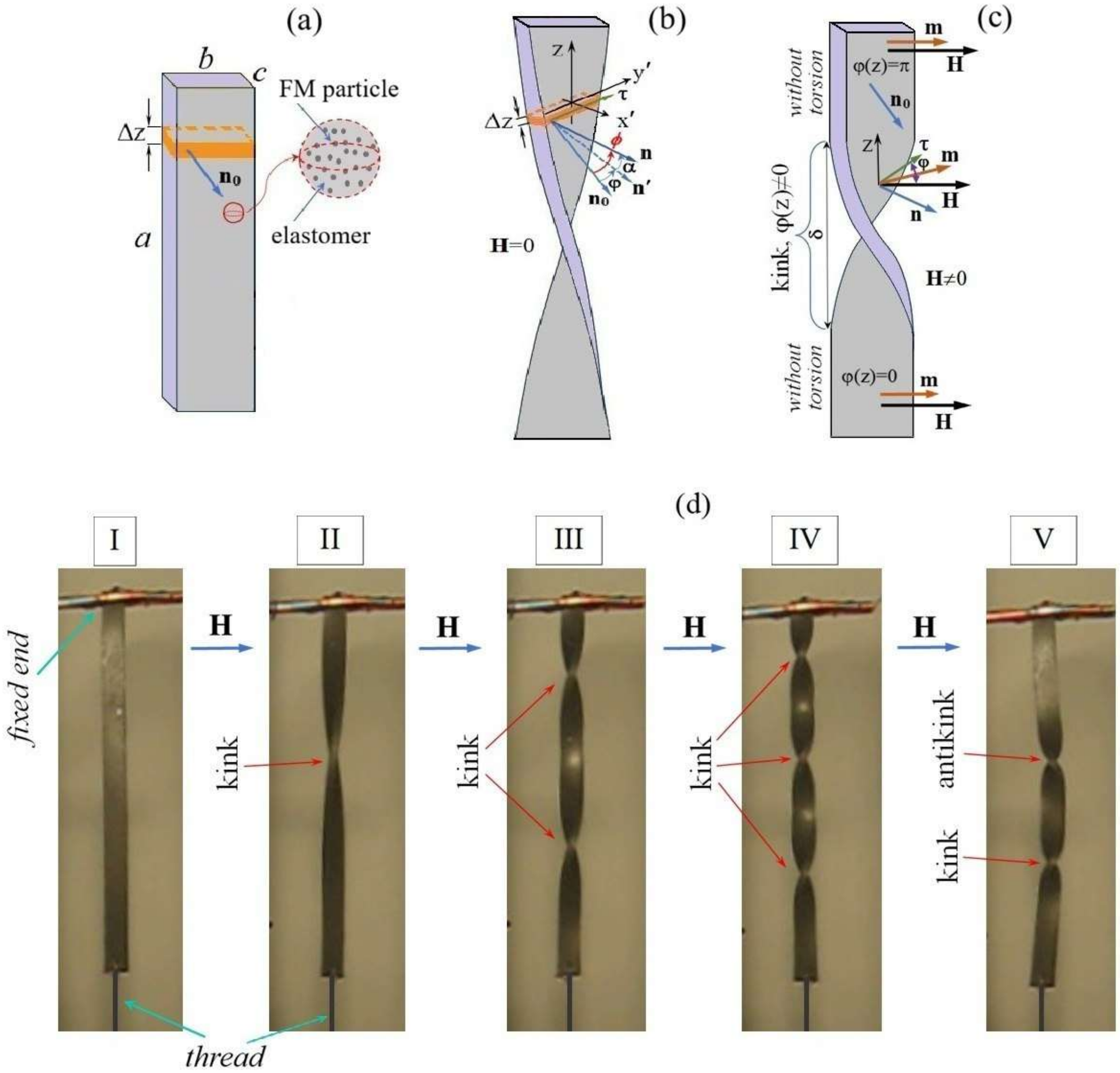

Figure 1. Image of the undeformed MAE beam (a); the beam with uniform torsion in the absence of a magnetic field, $H$=0, (b); image of the beam with a single kink (c); experimental photos obtained for the undeformed beam, the beam with one, two, and three kinks, and for a kink + anti-kink pair (d).

Figure 1(b) displays an image of this beam with uniform torsion deformation in the absence of a magnetic field ($H$=0). Unlike other materials, owing to the high elasticity of MAEs, the torsional deformation of the beam is reversible even for large twisting angles. When external moments are removed, the MAE beam recovers its initial undeformed state.

For a thin, long beam undergoing torsion, the St. Venant approximation[38] holds regarding the non-deforming flat cross-sections of the beam perpendicular to its long axis. For uniform torsion of a non-magnetized beam ($H$=0) by an angle $\pi$, the derivative of the angle of rotation for small sections of the beam of thickness $\Delta z$ (Figures 1(a) and 1(b)) is constant: $\partial\varphi/\partial z = \pi/a$, with torsion angles at the ends $\varphi(z = a/2) = \pi$ and $\varphi(z = -a/2) = 0$.

If a uniformly twisted beam is placed in a uniform magnetic field parallel to the beam plane and directed along its width, **H**||b, the influence of the moment of forces created by the magnetic field causes the torsion region to localize, forming a kink-like torsion (Figure 1(c)), whose length is denoted by $\delta$. Outside the kink region, there is no torsion; this section remains in a high-symmetry state with magnetization parallel to the magnetic field (**m**||**H**), where the external magnetic field does not generate a torsional moment of forces. Outside the kink region, the beam is unstressed, and no external moments are applied to the beam ends. In the kink region, the magnetization is non-collinear with the magnetic field; the beam's torsion in this region is solely due to the counteracting elastic and magnetoelastic moments.

In the St. Venant approximation, the linear density of the elastic energy in the kink region can be written as: $w_e(z) = GJ(\partial\varphi/\partial z)^2/2$, where $G$ is the shear modulus, and $J$ is the St. Venant constant. Its value is $J = kbc^3$ or $J = kc^2S$ where $k$ is a coefficient (for a very thin beam, $c/b\rightarrow 0$, $k$=1/3), and $S$ is the cross-sectional area of the beam, $S$=$bc$. For a beam of finite thickness, $k$ is less than 1/3.

During the beam's torsion, its initial normal vector rotates around the $z$-axis by an angle $\varphi$, transforming into the vector $\mathbf{n}'$ (Figure 1(b)). Due to a rotation around the $y'$-axis by an angle $\alpha$ (Figure 1(b)), the vector $\mathbf{n}'$ transforms into the normal vector **n** of the deformed beam. The angle between vectors $\mathbf{n}_0$ and **n** is denoted $\phi$.

For elementary rotation angles, the conditions $d\phi=d\varphi+d\alpha$ and $d\varphi\perp d\alpha$ hold. Using these, we obtain $\partial\phi/\partial z = \sqrt{(\partial\varphi/\partial z)^2 + (y'\partial^2\varphi/\partial z^2)^2}$. If the radius of curvature of the deformed beam surface along the $z$-axis is less than $\delta$, or the second term under the root is small enough to be neglected, we can assume the normal vector in the kink region is approximately perpendicular to the beam axis, $\mathbf{n}\perp z$ [38].

The rotation of the normal vector will be equal to the angle between the magnetic field strength vector **H** and the unit vector $\tau$, which is directed along $y'$. When the field strength vector **H** making an angle $\varphi$ (Figure 1(c)) with the beam plane, the magnetization projection onto the beam plane (onto the τ vector) will be $m_{\|} = \chi_{\|}\mu_0 H\cos\varphi$, and the magnetization projection onto the normal to the plane (onto the **n** vector) will be $m_{\perp} = \chi_{\perp}\mu_0 H\sin\varphi$, where $\mu_0$ is magnetic constant, $\chi_{\|}$, and $\chi_{\perp}$ are the longitudinal and transverse components of the MAE beam's magnetic susceptibility, with $\chi_{\|} > \chi_{\perp}$.

Accordingly, the linear density of the magnetic energy along *z* will be:

$$w_m(z) = -S\left(\chi_{\|}\mu_0 H^2\cos^2\varphi(z) + \chi_{\perp}\mu_0 H^2\sin^2\varphi(z)\right)/2. \quad (1)$$

The energy density $w_m$ depends on the torsion angle and the magnetic field strength; it describes the magnetoelastic contribution to the energy of the magnetized beam. The total energy of the beam's torsion kink can be written as:

$$W = \int_{-\infty}^{+\infty}\left(w_e(z) + w_m(z)\right)dz = -\frac{aS}{2}\chi_{\perp}\mu_0 H^2 + \frac{1}{2}\int_{-\infty}^{+\infty}\left(GJ\left(\frac{\partial\varphi}{\partial z}\right)^2 - S\Delta\chi\mu_0 H^2\cos^2\varphi\right)dz, \quad (2)$$

where $\Delta\chi = \chi_{\|} - \chi_{\perp}$, and the integration limits are $\pm\infty$, assuming the longitudinal size of the kink is smaller than the sample length.

Minimizing the energy (3) leads to the Euler-Lagrange differential equation in the form:

$$GJ\frac{\partial^2\varphi}{\partial z^2} - S\Delta\chi\mu_0 H^2\cos\varphi\sin\varphi = 0 \quad (3)$$

with boundary conditions $\partial\varphi/\partial z\big|_{z=\pm\infty} = 0$. The first term in (3) represents the linear density of the elastic moment of forces arising from the beam torsion. The second term is the density of the magnetoelastic moment of forces, whose magnitude depends on the magnetic field strength and the torsion angle. Equation (3) describes the equilibrium state: the elastic moment of forces is precisely compensated by the magnetoelastic moment of forces. Note that Equation (3) does not include any other external moments of applied forces.

The solution to Equation (3) for the case of a kink-like beam torsion, independent of its length, takes the form:

$$\varphi(z) = 2\arctan e^{z/\lambda}, \text{ where } \lambda = \sqrt{\frac{GJ}{S\Delta\chi\mu_0 H^2}}. \quad (4)$$

Thus, kink-like torsion can be stabilized in the beam by a magnetic field if the kink length is smaller than the beam length, $\delta \approx \pi\lambda < a$.

The expression for $\lambda$ can be further transformed using the geometric dimensions of the beam. For a very thin beam ($c/b \to 0$), we obtain:

$$\lambda = c\sqrt{\frac{G}{3\Delta\chi\mu_0 H^2}}. \quad (5)$$

The term under the root essentially represents the ratio of numerator and denominator having the dimensionality of volume energy density. It follows from expression (4) that a stronger magnetic field *H* results in a smaller kink size. Expression (5) for $\lambda$ is independent of the beam length, although the kink length increases with the beam width. This is because the wider the beam, the greater its resistance to torsion due to the increased lever arm of the elastic force; the parameter $J$ is proportional to the product of the square of the beam width and its cross-sectional area $S$. The elastic moment depends on the shear modulus $G$, whose magnitude can be small for MAE samples. Consequently, even a weak magnetic field is capable of stabilizing the kink, ensuring its size remains smaller than the beam length.

The magnetic anisotropy of the MAE beam's magnetization is a necessary condition for the kink to appear; it follows from (5) that the larger the difference $\Delta\chi > 0$, the smaller the kink length will be. In essence, Equation (3) describes the rotation along the $z$-axis of the beam's centrelines, in Fig. 1(b), the centreline coincides with the coordinate axis $y'$.

To confirm the formation of torsion kinks, we performed an experiment according to the scheme shown in Figs. 1 (a)–(c). We used a DC electromagnet with a pole diameter of $1.65\times10^{-1}$ m and a pole gap of $8.2\times10^{-2}$ m. This setup ensures high homogeneity of the magnetic field at the sample location. The dimensions of the sample were $a = 7.3\times10^{-2}$ m, $b = 4.7\times10^{-3}$ m, and $c = 2.1\times10^{-3}$ m. The sample was positioned midway between the magnetic poles, oriented perpendicularly to the magnetic field.

The sample fabrication process involved two stages[8,21,28]: (a) mixing and (b) curing. A two-component soft silicone compound, Silikon SKL-003 (Slovakia), with a density of $1.1\times10^3$ kg/m$^3$, a hardness of 3 Shore A units, and an elongation at break of 500 %, was utilized as the matrix material. The MAE sample contains 50 wt % of carbonyl iron particles embedded into the matrix. The particles are nearly spherical, with an average size of 3.3 μm determined via optical image processing. The spatial distribution of the particles within the matrix is uniformly random.

Magnetic properties were measured using an LDJ-9500 vibrating sample magnetometer. The saturation magnetization of the carbonyl iron powder is 205 A·m$^2$/kg. The saturation magnetization of the MAE containing carbonyl iron particles within the silicone matrix is 98.5 A·m$^2$/kg, and the magnetic susceptibility difference for the sample is $\Delta\chi = 0.25$. The sample is magnetically soft, exhibiting a coercive force of 1.2 kA/m and a remanent magnetization of 0.45 A·m$^2$/kg.

The shear modulus of the MAE is $G$=16 kPa. According to Ref. 39, such an MAE should be classified as elastically soft. Thus, the synthesized MAE sample represents a highly elastic ferromagnetic composite material that undergoes mechanical deformation easily and is highly sensitive to an external magnetic field. During the deformation and magnetization of an elastically soft MAE, displacements and rotations of the particles relative to the matrix can occur[13,40], which may lead to hysteretic phenomena[8,17,28] even in the case of magnetically soft filler particles. In our experiment, complete reversibility is observed: upon removing the magnetic field, the sample fully recovered its initial shape. The formation of the kink-like torsion deformation has a magnetoelastic nature and is not associated with the material's rheology.

Initially, in the absence of an external magnetic field ($H$=0), a mechanical torque was applied to the lower end of the beam, creating uniform torsion in the beam by an angle of $\pi$, $2\pi$, $3\pi$. Subsequently, a magnetic field was introduced, lying in the plane of the beam and perpendicular to its axis. Under the action of a finite magnetic field ($H\neq0$), the effect of torsion deformation localization was observed in the beam. After the kink/kinks were formed, the lower end of the beam was no longer subjected to external mechanical torsion, and the thread only prevented the beam from bending deformation.

Figure 1(d) shows the experimentally obtained images of the MAE beam with a single kink (photo II), with two kinks twisting in the same direction (photo III), with three kinks also twisting in the same direction (photo IV), and a kink + anti-kink system with mutually opposite directions of torsion (photo V). In the experiment, the upper end of the beam was rigidly fixed, preventing its rotation. The lower end of the beam, at its bottom point on the beam axis, was attached to a vertically oriented, practically slack thread. The thread's torsion around its axis requires negligible effort, and any possible longitudinal force exerted along the beam during torsion was negligibly small. Thus, the lower end of the MAE beam with kink(s) could freely rotate around its axis. Nonetheless, the beam states shown in photos II–V were stable in the experiment and did not spontaneously untwist unless the magnetic field strength was decreased. In a constant uniform magnetic field, the kink size remained unchanged over time. As the magnetic field increased, the kink size decreased, but no new/additional kinks were formed. Upon decreasing the magnetic field strength, the three-kink state abruptly transitioned to the two-kink state at a field $H_3$=59 kA/m. At a field $H_2$=39 kA/m, the two-kink state abruptly transitioned to the single-kink state. The transition from the single-kink state to the homogeneous beam state occurred at a field $H_1$=19 kA/m. For the studied MAE beam, these fields represent the lowest fields (critical fields) of stability for the beam states containing torsion kinks. When the magnetic field is removed, the kink and anti-kink disappear abruptly and simultaneously.

Using Eq. (5), the kink size can be estimated at the critical field $H_1$, yielding $\lambda$=1.5×10$^{-2}$ m; as expected, the characteristic size of the kink at $H$=$H_1$ is comparable to the beam length, $\delta\sim a$. At the critical field $H_2$, the kink size is slightly less than half the beam length, and for three kinks at the critical field $H_3$, the kink size is slightly less than one-third of the beam length. It can be assumed that the magnetic field-stabilized torsion kink loses its stability when its size equals the beam length. In the case of multiple kinks, stability is lost in a magnetic field where the kink length is equal to the beam length divided by the number of kinks. Therefore, the magnitude of the critical stability field for a beam with multiple kinks is inversely proportional to the number of kinks.

In conclusion, we have theoretically and experimentally investigated a novel effect of kink-like deformation formation in an MAE beam subjected to a permanent uniform magnetic field. The MAE beam with a torsion kink consists of two uniformly magnetized, undeformed regions separated by a non-uniformly deformed MAE section in a low-symmetry magnetic state. The energy loss in the localized, highly non-uniform kink section is compensated by the gain in magnetic energy in the uniform sections of the beam. The formation of the kink is driven by volumetrically distributed and mutually compensating elastic moments of forces and magnetoelastic moments of forces within the deformed beam. The magnetic field-induced formation of this equilibrium structure in the MAE beam is well confirmed by the experiment where one end of the beam is free to rotate.

The research described in this paper was partially supported by the IEEE Magnetics Society Program “Magnetism for Ukraine 2024/2025” (STCU Project No. 9918).

## AUTHOR DECLARATIONS

### Conflict of Interest

The authors have no conflicts to disclose.

### Author Contributions

**Yu. I. Dzhezherya** https://orcid.org/0000-0003-1560-9567

Writing – original draft (equal); Supervision (equal); Project administration (lead); Formal analysis (equal); Investigation (equal).

**A. V. Kyryliuk** https://orcid.org/0009-0002-7735-8804

Conceptualization (equal); Data curation (equal); Formal analysis (equal); Investigation (equal).

**S. V. Cherepov** https://orcid.org/0000-0001-7257-2911

Conceptualization (equal); Data curation (equal); Investigation (equal).

**Yu. B. Skirta** https://orcid.org/0009-0008-4656-9450

Conceptualization (equal); Data curation (equal); Investigation (equal).

**S. O. Reshetniak** https://orcid.org/0000-0003-2316-5396

Formal analysis (equal); Writing – original draft (equal); Conceptualization (equal).

**S. M. Ryabchenko** https://orcid.org/0000-0002-5638-7166

Methodology (equal); Conceptualization (equal); Formal analysis (equal).

**V. M. Kalita** https://orcid.org/0000-0001-6329-9095

Formal analysis (equal); Supervision (equal); Writing – original draft (equal); Writing – review & editing (lead).

## DATA AVAILABILITY

The data that support the findings of this study are available from the corresponding authors upon reasonable request.